\documentclass[twocolumn,pra]{revtex4}
\usepackage{graphicx}

\begin{document}

%\preprint{}

\title{Atomic Clocks and Constraints on Variations of Fundamental
Constants}

\author{Savely G. Karshenboim}
\affiliation{D. I. Mendeleev Institute for Metrology (VNIIM), St. Petersburg 198005, Russia\\
Max-Planck-Institut f\"ur Quantenoptik, 85748 Garching, Germany}
\author{Victor Flambaum}
\affiliation{School of Physics, University of New South Wales, Sydney 2052, Australia}
\author{Ekkehard Peik} 
\affiliation{Physikalisch-Technische Bundesanstalt, 38116 Braunschweig, Germany}

\date{}

\maketitle

\label{cxx-99}
\section{Introduction}%xx.1
\label{sec:intro-99}
\index{Fundamental~constants}

Fundamental constants play an important role in modern physics,
being landmarks that designate different areas. We call them
constants, however, as long as we only consider minor variations
with the cosmological time/space scale, their constancy is an
experimental fact rather than a basic theoretical principle.
Modern theories unifying gravity with electromagnetic, weak, and
strong interactions, or even the developing quantum gravity itself
often suggest such variations.

Many parameters that we call {\it fundamental constants\/}, such
as the electron charge and mass (see, e.g., Ref.\
\cite{here:Drake-99,codata-99}), are actually not truly
fundamental constants but effective parameters which are affected
by renormalization or the presence of matter \cite{sgk-99}. Living
in a changing universe we cannot expect that matter will affect
these parameters the same way during any given cosmological epoch.
An example is the inflationary model of the universe which states
that in a very early epoch the universe experienced a phase
transition which, in particular, changed a vacuum average of the
so-called Higgs field which determines the electron mass. The
latter was zero before this transition and reached a value close
or equal to the present value after the transition.

The problem of variations of constants has many facets and here we
discuss aspects related to atomic clocks and precision frequency
measurements. Other related topics may be found in, e.g., Ref.\
\cite{book-99}.

Laboratory searches for a possible time variation of fundamental
physical constants currently consist of two important parts: (i)
one has to measure a certain physical quantity at two different
moments of time that are separated by at least a few years; (ii)
one has to be able to interpret the result in terms of fundamental
constants. The latter is a strong requirement for a cross
comparison of different results.

The measurements which may be performed most accurately are
frequency measurements; and thus, frequency standards or atomic
clocks will be involved in most of the laboratory searches.
Frequency metrology has shown great progress in the last decade
and will continue to do so for some time. The constraints on the
variations of the fundamental constants obtained in this manner
are, so far, somewhat weaker than those from other methods
(astrophysics, geochemistry), but still competitive with them. In
contrast to other methods, however, frequency measurements allow a
very clear interpretation of the final results and a transparent
evaluation procedure, making them less vulnerable to systematic
errors. While there is still potential for improvement, the basic
details of the method have been recently fixed.

The most advanced atomic clocks are discussed in
Sect.~\ref{sec:ac-99}. They are realized with many-electron atoms
and their frequency cannot be interpreted in terms of fundamental
constants. However, a much simpler problem needs to be solved: to
interpret their variation in terms of fundamental constants. This
idea is discussed in Sect.~\ref{sec:ams-99}. The current
laboratory constraints on the variations of the fundamental
constants are summarized in Sect.~\ref{sec:clc-99}.

\section{Atomic clocks and frequency
 standards}%xx.2
\label{sec:ac-99}
\index{Atomic clocks}
\index{Frequency standard}

Frequency standards are important tools for precision measurements
and serve various purposes which, in turn, have different
requirements that must be satisfied. In particular, it is not
necessary for a frequency standard to reproduce a frequency which
is related to a certain atomic transition although it may be
expressed in its terms. A well known example is the hydrogen
maser, where the frequency is affected by the wall shift which may
vary with time \cite{ramsey-99}. For the study of time variations
of fundamental constants it is necessary to use standards similar
to a primary caesium clock. In this case, any deviation of its
frequency from the unperturbed atomic transition frequency should
be known (within a known uncertainty) because this is a necessary
requirement for being a `primary' standard.

From the point of view of fundamental physics, the hydrogen maser
is an artefact quite similar, in a sense, to the prototype of the
kilogram held at the Bureau International des Poids et Mesures
(BIPM) in Paris. Both artefacts are somehow related to fundamental
constants (e.g., the mass of the prototype can be expressed in
terms of the nucleon masses and their number) but they also have a
kind of residual classical-physics flexibility which allows their
properties to change. In contrast, standards similar to the
caesium clock have a frequency (or other property) that is
determined by a certain natural constant which is not flexible,
being of pure quantum origin. It may change only if the
fundamental constants are changing.

In Sect.~\ref{sec:clc-99}, results obtained with caesium and
rubidium fountains, a hydrogen beam, ultracold calcium clouds, and
trapped ions of ytterbium and mercury are discussed. While caesium
and rubidium clocks operate in the radio frequency domain, most of
the other standards listed above rely on optical transitions.

\subsection{Caesium Atomic Fountain}%
\label{susec:af-99}
\index{Atomic fountain}

\begin{figure}[htb]
%\begin{center}
\includegraphics[width=0.4\textwidth]{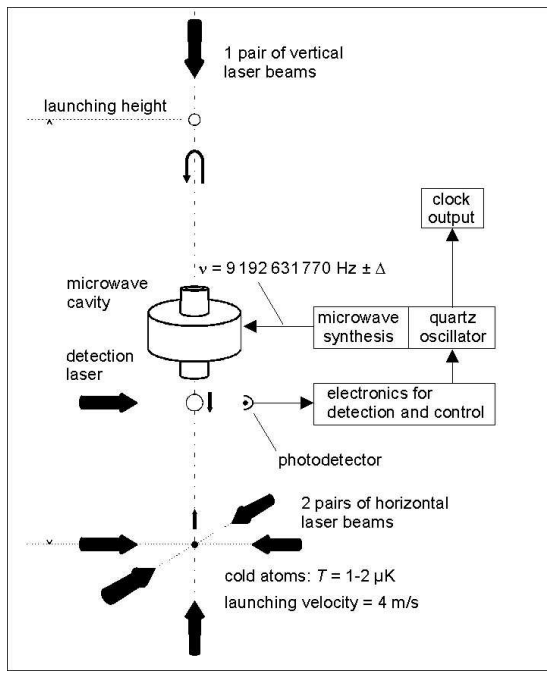}
%\epsfig{figure=fount.eps,width=65mm}
%\end{center}
\caption{Schematic of an atomic fountain clock.}
\label{fig1f-99}
\end{figure}

Caesium clocks are the most accurate primary standards for time
and frequency \cite{bauch-99}. The hyperfine splitting frequency
between the $F=3$ and $F=4$ levels of the $^2S_{1/2}$ ground state
of the $^{133}$Cs atom at 9.192~GHz has been used for the
definition of the SI second since 1967. In a so-called caesium
fountain (see Fig.~\ref{fig1f-99}), a dilute cloud of laser cooled
caesium atoms at a temperature of about 1~$\mu$K is launched
upwards to initiate a free parabolic flight with an apogee at
about 1~m above the cooling zone. A microwave cavity is mounted
near the lower endpoints of the parabola and is traversed by the
atoms twice -- once during ascent, once during descent -- so that
Ramsey's method of interrogation with separated oscillatory fields
\cite{ramsey-99} can be realized. The total interrogation time
being on the order of 0.5~s, a resonance linewidth of 1~Hz is
achieved, about a factor of 100 narrower than in traditional
devices using a thermal atomic beam from an oven. Selection and
detection of the hyperfine state is performed via optical pumping
and laser induced resonance fluorescence. In a carefully
controlled setup, a relative uncertainty slightly below\linebreak
 $1\cdot 10^{-15}$ can be reached in the
realization of the resonance frequency of the unperturbed Cs atom.
The averaging time that is required to reach this level of
uncertainty is on the order of $10^4$ s. One limiting effect that
contributes significantly to the systematic uncertainty of the
caesium fountain is the frequency shift due to cold collisions
between the atoms. In this respect, a fountain frequency standard
based on the ground state hyperfine frequency of the $^{87}$Rb
atom at about 6.835~GHz is more favorable, since its collisional
shift is lower by more than a factor of 50 for the same atomic
density. With the caesium frequency being fixed by definition in
the SI system, the $^{87}$Rb frequency is therefore presently the
most precisely measured atomic transition frequency \cite{rb-99}.

\subsection{Single-Ion Trap}%xx.2
\label{susec:sit-99}
\index{Single-ion trap}

An alternative to interrogating atoms in free flight, and a
possibility to obtain practically unlimited interaction time, is
to store them in a trap. Ions are well suited because they carry
electric charge and can be trapped in radio frequency ion traps
(Paul traps \cite{paul-99}) that provide confinement around a
field-free saddle point of an electric quadrupole potential. This
ensures that the internal level structure is only minimally
perturbed by the trap. Combined with laser cooling it is possible
to reach the so-called Lamb--Dicke regime where the linear Doppler
shift is eliminated. A single ion, trapped in an ultrahigh vacuum
is conceptually a very simple system that allows good control of
systematic frequency shifts \cite{deh-99}. The use of the much
higher, optical reference frequency allows one to obtain a
stability that is superior to microwave frequency standards,
although only a single ion is used to obtain a correction signal
for the reference oscillator.

A number of possible reference optical transitions with a natural
linewidth of the order of 1 Hz and below are available in
different ions, such as Yb$^+$ \cite{yb-99} and Hg$^+$
\cite{hg-99}. These ions possess a useful level system, where both
a dipole-allowed transition and a forbidden reference transition
of the optical clock can be driven with two different lasers from
the ground state. The dipole transition is used for laser cooling
and for the optical detection of the ion via its resonance
fluorescence. If a second laser excites the ion to the metastable
upper level of the reference transition, the fluorescence
disappears and every single excitation can thus be detected with
practically hundred percent efficiency as a dark period in the
fluorescence signal.

\begin{figure}[htb]
\includegraphics[width=0.4\textwidth]{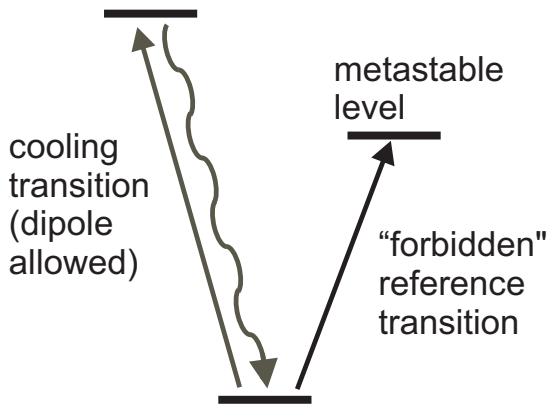}
%\centerline{\epsfig{figure=vlevel.eps,width=50mm}}
\caption{Double resonance scheme applied in  single-ion-trap frequency standards.}
\label{f:v-99}
\end{figure}

Using these techniques and a femtosecond laser frequency comb
generator (see Sect.~99.2.5) for the link to primary caesium
clocks, the absolute frequencies of the transitions
$^2S_{1/2}\rightarrow {^2 D_{5/2}}$ in $^{199}$Hg$^+$ at 1065 THz
and $^2S_{1/2}\rightarrow {^2 D_{3/2}}$ in $^{171}$Yb$^+$ at
688~THz have been measured with relative uncertainties of only
$9\cdot 10^{-15}$. It is believed that single-ion optical
frequency standards offer the potential to ultimately reach the
$10^{-18}$ level of relative accuracy.

A similar double resonance technique can be employed  if the
reference transition is in the microwave domain and a number of
accurate measurements of hyperfine structure intervals in trapped
ions has been performed. In particular, the HFS interval in
$^{171}$Yb$^+$ has been measured several times \cite{ybhfs-99} and
can be used to obtain constraints on temporal variations.

\subsection{Laser-Cooled Neutral Atoms}
\label{susec:mot-99}
\index{Magneto-optical trap}

Optical frequency standards have been developed with free
laser-cooled neutral atoms, most notably of the alkaline-earth
elements that possess narrow intercombination transitions. The
atoms are collected in a magneto-optical trap, are then released
and interogated by a sequence of laser pulses to realize a
frequency-sensitive Ramsey-Bord\'e atom interferometer
\cite{borde-99}. Of these systems, the one based on the
$^1S_0\rightarrow {^3P_1}$ intercombination line of $^{40}$Ca at
657 nm has reached the lowest relative uncertainty so far (about
$2\cdot 10^{-14}$) \cite{hg-99,ca-99}. Limiting factors in the
uncertainty of these standards are the residual linear Doppler
effect and phase front curvature of the laser beams that excite
the ballistically expanding atom cloud. It has therefore been
proposed to confine the atoms in an optical lattice, i.e., in the
array of interference maxima produced by several intersecting,
red-detuned laser beams \cite{katori-99}. The detuning of the
trapping laser could be chosen such that the light shift it
produces in the ground and excited state of the reference
transition are equal, and therefore it would produce no shift of
the reference frequency. This approach is presently being
investigated and may be applied to the very narrow (mHz natural
linewidth) $^1S_0\rightarrow {^3P_0}$ transitions in neutral
strontium, ytterbium, or mercury.

\subsection{Two-Photon Transitions and Doppler-Free Spectroscopy}%
\label{susec:dfs-99}
\index{Doppler-free spectroscopy}

The linear Doppler shift of an absorption resonance can also be
avoided if a two-photon excitation is induced by two
counterpropagating laser beams. A prominent example that has been
studied with high precision is the two-photon excitation of the
$1S\rightarrow 2S$ transition in atomic hydrogen. The precise
measurement of this frequency is of importance for the
determination of the Rydberg constant and as a test of quantum
electrodynamics (QED). Hydrogen atoms are cooled by collisions in
a cryogenic nozzle and interact with a standing laser-wave of
243~nm wavelength inside a resonator. Since the atoms are not as
cold as in laser cooled samples, a correction for the second order
Doppler effect is performed. The laser excitation is interrupted
periodically and the excited atoms are detected in a time resolved
manner so that their velocity can be examined. An accuracy of
about $2\cdot 10^{-14}$ has been obtained in absolute frequency
measurements with a transportable caesium fountain \cite{h-99}.

\subsection{Optical Frequency
Measurements}%
\label{susec:ofm-99}
\index{Frequency comb}

In recent years, the progress in stability and accuracy of optical
frequency standards has been impressive; and there is belief that
in the future an optical clock may supersede the microwave clocks
because the optical oscillators offer a much higher number of
periods in a given time. In addition, some systematic effects,
such as the Zeeman effect, have an absolute order of magnitude
that does not scale with the transition frequency, and
consequently is relatively less important at higher transition
frequencies. A long-standing problem, however, was the precise
conversion of an optical frequency to the microwave domain, where
frequencies can be counted electronically in order to establish a
time scale or can easily be compared in a phase coherent way.

\begin{figure}[htb]
\includegraphics[width=0.4\textwidth]{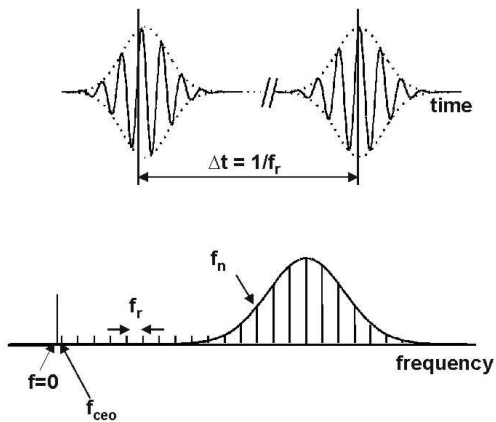}
%\centerline{\epsfig{figure=comb.eps,width=65mm}}
\caption{Frequency comb generated from femtosecond laser pulses.}
\label{fig1c-99}
\end{figure}

This problem has recently been solved by the so-called femtosecond
laser frequency comb generator \cite{chain-99}. Briefly, a
mode-locked femtosecond laser produces, in the frequency domain, a
comb of equally spaced optical frequencies $f_n$ that can be
written as $f_n=nf_r+f_{\rm ceo}$ (with $f_{\rm ceo}<f_r$), where
$f_r$ is the pulse repetition rate of the laser, the mode number
$n$ is a large integer (of order $10^5$), and $f_{\rm ceo}$
(carrier-envelope-offset) is a shift of the whole comb that is
produced by group velocity dispersion in the laser. The repetition
rate $f_r$ can easily be measured with a fast photodiode. In order
to determine $f_{\rm ceo}$, the comb is broadened in a nonlinear
medium so that it covers at least one octave. Now the second
harmonic of mode $n$ from the ``red'' wing of the spectrum, at
frequency $2(nf_r+f_{\rm ceo})$, can be mixed with mode $2n$ from
the ``blue'' wing, at frequency $2nf_r+f_{\rm ceo}$, and $f_{\rm
ceo}$ is obtained as a difference frequency. In this way, the
precise relation between the two microwave frequencies $f_r$ and
$f_{\rm ceo}$ and the numerous optical frequencies $f_n$ is known.
The setup can now be used for an absolute optical frequency
measurement by referencing $f_r$ and $f_{\rm ceo}$ to a microwave
standard and recording the beat note between the optical frequency
$f_{\rm o}$ to be measured and the closest comb frequency $f_n$.
Vice versa, the setup may work as an optical clockwork, for
example, by adjusting $f_{\rm ceo}$ to zero and by stabilizing one
comb line $f_n$ to $f_{\rm o}$ so that $f_r$ is now an exact
subharmonic to order $n$ of $f_o$. The precision of these transfer
schemes has been investigated and was found to be so high that it
will not limit the performance of optical clocks for the
foreseeable future.

\subsection{Limitations on Frequency Variations}%
\label{susec:lfv-99}

The frequency standards described above have been succesfully
developed and their accuracy has been improved in the last decade.
This progress, as a consequence, has led to certain constraints on
the possible variations of the fundamental constants. Considering
frequency variations, one has to have in mind that not only the
numerical value but also the units may vary. For this reason, one
needs to deal with dimensionless quantities which are
unit-independent. During the last decade, a number of transition
frequencies were measured in the corresponding SI unit, the hertz.
These dimensional results are actually related to dimensionless
quantities since a frequency measurement in SI is a measurement
with respect to the caesium hyperfine interval\footnote{Most
absolute frequency measurements have been realized as a direct
comparison with a primary caesium standard.}
\begin{equation}\label{fcs-99}
\bigl\{f\bigr\}= 9\,192\,631\,770 \; \mbox{}\cdot \frac{f}{f_{\rm
HFS}({\rm Cs})} \;,
\end{equation}
where $\bigl\{f\bigr\}$ stands for the numerical value of the
frequency $f$. In Sect.~\ref{sec:clc-99}, in order to simplify
notation, this symbol for the numerical value is dropped.

\begin{table}
\caption{Limits on possible time variation of frequencies of different
transitions in SI units. Here $\delta f/f$ is the fractional uncertainty
of the most accurate measurement of the frequency $f$.
\label{t:dfdt-99}}
\begin{center}
\begin{tabular*}{246pt}{rcccc}
\hline
Atom, & $f$ & $\delta f/f$  & $\Delta f/\Delta t$  & Refs. \\
transition & [GHz] & $[10^{-15}]$ & [Hz/yr] &  \\
\hline
H, Opt &2\,466\,061 &$14$ & $-8\pm 16$& \protect\cite{h-99}\\
Ca, Opt &455\,986 &$13$&$-4\pm5$& \protect\cite{ca-99}\\
Rb, HFS & 6.835&$1$&$(0\pm5)\cdot10^{-6}$& \protect\cite{rb-99}\\
Yb$^+$, Opt & 688\,359 &$9$ &$-1\pm3$& \protect\cite{ybnew-99}\\
Yb$^+$, HFS & 12.642 &$73$&$(4\pm4)\cdot10^{-4}$& \protect\cite{ybhfs-99}\\%0.35\pm0.37 mHz/yr;2.77+-2.94x10-14
Hg$^+$, Opt & 1\,064\,721 & $9$ &$0\pm7$& \protect\cite{hg-99}\\
\hline
\end{tabular*}
\end{center}
\end{table}

\section{Atomic spectra and % their dependence on
 the fundamental constants}%xx.3
\label{sec:ams-99}

\subsection{The Spectrum of Hydrogen and Nonrelativistic
Atoms}%xx.2
\label{susec:sha-99}
\index{Hydrogen atom}

The hydrogen atom is the simplest atom and one can easily
calculate the leading contribution to different kinds of
transitions in its spectrum ({\it cf.\/}, for example, Ref.\
\cite{here:Sapirstein-99}), such as the gross, fine, and hyperfine
structure. The scaling behavior of these contributions with the
values of the Rydberg constant $R_\infty$, the fine structure
constant $\alpha$, and the magnetic moments of proton and Bohr
magneton is clear. The results for some typical hydrogenic
transitions are
\begin{eqnarray}\label{eq:scal-99}
f(2p\to 1s)&\simeq&\frac{3}{4}\cdot cR_\infty\;,\nonumber\\
f(2p_{3/2}-2p_{1/2})&\simeq&\frac{1}{16}\cdot\alpha^2\cdot
cR_\infty\;,\nonumber\\
f_{\rm HFS}(1s) &\simeq&\frac{4}{3}\cdot\alpha^2\cdot
\frac{\mu_{\rm p}}{\mu_{\rm B}}
 \cdot cR_\infty\;.
%\mu \cdot \frac{m_e}{m_p}
\end{eqnarray}

In the nonrelativistic approximation, the basic frequencies and
the fine and hyperfine structure intervals of all atomic spectra
have a similar dependence on the fundamental constants. The
presence of a few electrons and a nuclear charge of $Z\neq1$ makes
theory more complicated and introduces certain multiplicative
numbers but involves no new parameters. The importance of this
scaling for a search for the variations was first pointed out in
Ref.\ \cite{savedoff-99} and was applied to astrophysical data.
Similar results may be presented for molecular transitions
(electronic, vibrational, rotational and hyperfine)
\cite{thompson-99}, however, up to now no measurement with
molecules has been performed at a level of accuracy that is
competitive with atomic transitions. They have been used only in a
search for variations of constants in astrophysical observations
(see e.g. \cite{varshalovich-99}).

\subsection{Hyperfine Structure and the Schmidt Model}%xx.2
\label{susec:hfs-99}
\index{Schmidt model}

The atomic hyperfine structure
\begin{equation}\label{NRHFS-99}
f_{\rm NR}({\rm HFS}) = {\rm const} \cdot \alpha ^2 \cdot
\frac{\mu}{\mu_{\rm B}} \cdot cR_\infty
\end{equation}
involves nuclear magnetic moments $\mu$ which are different for
different nuclei; thus, a comparison of the constraints on the
variations of nuclear magnetic moments has a reduced value. To
compare them, one may apply the Schmidt model (see, e.g., Ref.\
\cite{Karshenboim-99,sgk-99}), which predicts all the magnetic
moments of nuclei with an odd number of nucleons (odd value of
atomic number $A$) in terms of the proton and neutron $g$-factors,
$g_{\rm p}$ and $g_{\rm n}$, respectively, and the nuclear
magneton only. Unfortunately, the uncertainty of the calculation
within the Schmidt model is quite high (usually from 10\% to
50\%).  The Schmidt model, being a kind of {\em ab initio\/}
model, only allows for improvements which, unfortunately, involve
some effective phenomenological parameters. This would not really
improve the situation, but return us to the case where there are
too many possibly varying independent parameters. A comparison of
the Schmidt values to the actual data is presented for caesium,
rubidium, and ytterbium in Table~\ref{t:hfs-99}.

\begin{table}
\caption{Magnetic moments and relativistic corrections for atoms
involved in microwave standards. The relativistic sensitivity
$\kappa$ is defined in Sect.~\ref{susec:asrc-99}. Here $\mu$ is an
actual value of the nuclear magnetic moment, $\mu_{\rm N}$ is the
nuclear magneton, and $\mu_{\rm S}$ stands for the Schmidt value
of the nuclear magnetic moment; the nucleon $g$ factors are
$g_{\rm p}/2\simeq 2.79$ and $g_{\rm n}/2\simeq-1.91$.
\label{t:hfs-99}}
\begin{center}
\begin{tabular*}{246pt}{cccccc}
\hline
$Z$&Atom & $\mu/\mu_{\rm N}$  & $\mu_{\rm S}/\mu_{\rm N}$ &
$\mu/\mu_{\rm S}$ &
$\kappa$ \\
\hline
37  &$^{87}$Rb & 2.75 &$g_{\rm p}/2+1$ &0.74& 0.34 \\
55&$^{133}$Cs   & 2.58 &$7/18\cdot(10-g_{\rm p})$&1.50& 0.83\\
70&$^{171}$Yb$^+$ & 0.49 &$-g_{\rm n}/6$&0.77& 1.5 \\
\hline
\end{tabular*}
\end{center}
\end{table}

\subsection{Atomic Spectra: Relativistic Corrections}%xx.2
\label{susec:asrc-99}
\index{Relativistic corrections}

A theory based on the leading nonrelativistic approximation may
not be accurate enough. Any atomic frequency can be presented as
\begin{equation}
f = f_{\rm NR}\cdot F_{\rm rel}(\alpha)\;,
\end{equation} where the first (nonrelativistic) factor is
determined by a scaling similar to the hydrogenic transitions
(\ref{eq:scal-99}). The second factor stands for relativistic
corrections which vanish at $\alpha=0$; and thus, $F_{\rm
rel}(0)=1$.

The importance of relativistic corrections for the hyperfine
structure was first emphasized in Ref.\ \cite{prestage-99}.
Relativistic many-body calculations for various transitions have
been performed in Refs.\
\cite{flambaum04-99,dzuba1-99,dzuba01-99,dzuba03-99}. A typical
accuracy is about 10\%. Some results are summarized in
Tables~\ref{t:hfs-99} and~\ref{t:opt-99}, where we list the
relative sensitivity of the relativitic factors $F_{\rm rel}$ to
changes in $\alpha$,
\begin{equation}\label{kap-99}
\kappa = \frac{\partial \ln F_{\rm rel}}{\partial \ln \alpha}\;.
\end{equation}
Note that the relativistic corrections in heavy atoms are
proportional to $(Z \alpha)^2$ because of the singularity of
relativistic operators. Due to this, the corrections rapidly
increase with the nuclear charge $Z$.

The signs and magnitudes of  $\kappa$ are explained by a simple
estimate of the relativistic correction. For example, an
approximate expression for the relativistic correction factor for
the hyperfine structure of an $s$-wave electron in an alkali-like
atom is (see, e.g., Ref.\ \cite{prestage-99})
\[
F_{\rm rel}(\alpha) =
\frac{1}{\sqrt{1-(Z \alpha)^2}}
\cdot
\frac{1}{1-(4/3)(Z\alpha)^2}\simeq 1+\frac{11}{6}(Z\alpha)^2\;.
\]
A similar rough estimation for the energy levels may be performed
for the gross structure:
\begin{equation}\label{rel-99}
E = -\frac{Z_{a}^{2}mc^{2}\alpha^{2}}{2n_*^{2}}\cdot\left(1+
\frac{(Z\alpha)^{2}}{n_*}\frac{1}{j + 1/2}\right) \;.
\end{equation}
Here $j$ is the electron angular momentum, $n_*$ is the effective
value of the principle quantum number (which determines the
nonrelativistic energy of the electron), and $Z_{a}$ is the charge
``seen'' by the valence electron -- it is 1 for neutral atoms, 2
for singly charged ions, etc. This equation tells us that
$\kappa$, for the excitation of the electron from the orbital $j$
to the orbital $j^\prime$, has a different sign for $j>j^\prime$
and $j<j^\prime$. The difference of sign between the sensitivities
of the ytterbium and mercury transitions in Table~\ref{t:opt-99}
reflects the fact that in Yb$^+$ a $6s$-electron is excited to the
empty $5d$-shell, while in Hg$^+$ a hole is created in the filled
$5d$-shell if the electron is excited to the $6s$-shell.

\begin{table}
\caption{Limits on possible time variation of the frequencies of
different transitions and their sensitivity to variations in
$\alpha$ due to relativistic corrections. \label{t:opt-99}}
\begin{center}
\begin{tabular*}{246pt}{lrr@{}l}
\hline
\multicolumn{1}{c}{Atom, transition} & \multicolumn{1}{c}{$\partial f/f \partial t$} & \multicolumn{2}{c}{$\kappa$} \\
\hline
H, $1s-2s$ & $-3.2(63)\times10^{-15}$ yr$^{-1}$ & 0.&00\\
$^{40}$Ca,  ${^1}S_0-^3P_1$ &$-8(11)\times10^{-15}$ yr$^{-1}$ &0.&03 \\ %-7.9(110)
$^{171}{\rm Yb}^+$, ${^2}S_{1/2}-{^2}D_{3/2}$ & $-1.2(44)\times10^{-15}$ yr$^{-1}$  & 0.&9\\
%$^{171}{\rm Yb}^+$, ${^2}S_{1/2}-{^2}D_{3/2}$ & $(-1.2\pm 4.4)\times10^{-15}$ yr$^{-1}$  & 0.&9\\
$^{199}{\rm Hg}^+$, $^2S_{1/2}-{^2}D_{5/2}$ & $-0.2(70)\times10^{-15}$ yr$^{-1}$ & $-3.$&2\\
\hline
\end{tabular*}
\end{center}
\end{table}

\section{%Current l
Laboratory constraints on % time
the variations of the fundamental constants}%xx.4
\label{sec:clc-99}
\index{Fundamental~constants}

Logarithmic derivatives [see, e.g., Eq.\ (\ref{kap-99})] appear
since we are looking for a variation of the constants in relative
units. In other words, we are interested in a determination of,
e.g., $\Delta \alpha/\alpha\Delta t$ while the input data of
interest are related to $\Delta f/f\Delta t$. Their relation takes
the form
\begin{equation}\label{eq1-99}
\frac{\partial \ln f}{\partial t} = \frac{\partial \ln f_{\rm
NR}}{\partial t} + \kappa\cdot \frac{\partial \ln \alpha}{\partial
t}\;.
\end{equation}
If one compares transitions of the same type  -- gross structure,
fine structure -- the first term cancels.

\subsection{Constraints from Absolute Optical Measurements}%xx.2
\label{susec:aom-99}

\begin{figure}[htb]
\includegraphics[width=0.4\textwidth]{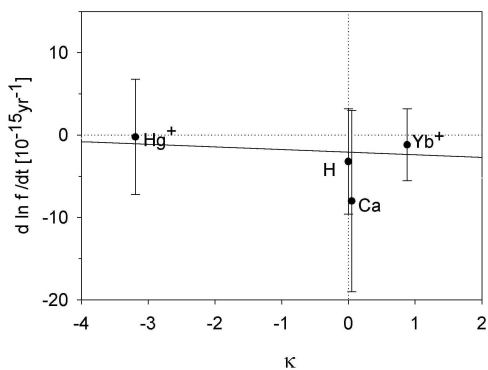}
%\centerline{\epsfig{figure=linreg.eps,width=75mm}}
\caption{Frequency variations versus their sensitivity $\kappa$.}
\label{f:const-99}
\end{figure}

Absolute frequency measurements offer the possibility to compare a
number of optical transitions with frequencies $f_{\rm NR}$, which
scale as $cR_\infty$, with the caesium hyperfine structure. One
can rewrite Eq.~(\ref{eq1-99}) as
\begin{equation}\label{eqopt-99}
\frac{\partial \ln f_{\rm opt}}{\partial t} = \frac{\partial \ln
cR_\infty}{\partial t} + \kappa\cdot \frac{\partial \ln
\alpha}{\partial t}\;,
\end{equation}
where dimensional quantities, such as frequency and the Rydberg
constant, are stated in SI units [{\it cf.\/} Eq.~(\ref{fcs-99})].
This equation may be used in different ways. For example, in
Fig.~\ref{f:const-99} we plot experimental data for $\partial \ln
f_{\rm opt}/\partial t$ as a function of the sensitivity $\kappa$
and derive a model-independent constraint on the variation of the
fine structure constant
\begin{equation}\label{eqalp-99}
 \frac{\partial \ln \alpha}{\partial t} = (-0.3\pm 2.0)\cdot 10^{-15}\,{\rm yr}^{-1}
\end{equation}
and the numerical value of the Rydberg frequency $cR_\infty$ (see
Table~\ref{t:mic-99}) in the SI unit of hertz. The latter is of
great metrological importance, being related to a common drift of
optical clocks with respect to a caesium clock, i.e., to the
definition of the SI second. The SI definition of the metre is
unpractical and so, in practice, the optical wavelengths of
reference lines calibrated against the caesium standard are used
to determine the SI metre \cite{quinn-99}.

\begin{figure}[htb]
%\centerline{\epsfig{figure=2dplot.eps,width=75mm}}
\includegraphics[width=0.4\textwidth]{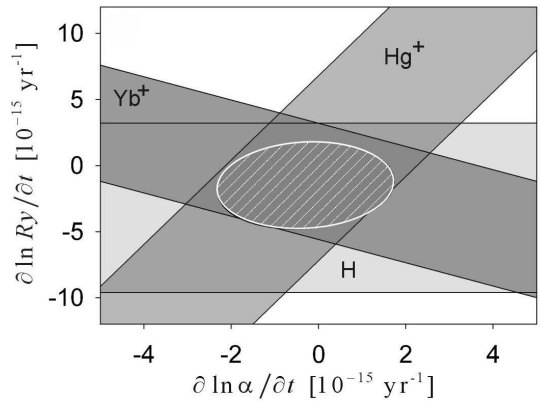}
\caption{Constraints on the time variations of the fine structure
constant $\alpha$ and the numerical value of the Rydberg constant.
The preliminary data on Ca are not included.} \label{fig1a-99}
\end{figure}

The constraints on the variations of $\alpha$ and $cR_\infty$ are
correlated and the standard uncertainty ellipse, defined as
%\begin{equation}%\label{ellipse-99}
\[
\sum_{i} \frac{1}{u_i^2}\left( \frac{\partial \ln f_i}{\partial
t}-\frac{\partial \ln Ry}{\partial t}- \kappa_i\frac{\partial \ln
\alpha}{\partial t}\right) ^2 =1+\chi^2_{\rm min} \, ,
\]
%\end{equation}
is presented in Fig.~\ref{fig1a-99}. Here we sum over all
available data: $\partial \ln f_i/\partial t$ is the central value
of the observed drift rate, $u_i$ its $1\sigma$ uncertainty, and
$\chi^2_{\rm min}$ the minimized $\chi^2$ of the fit.

The numerical value of the Rydberg constant, from the point of
view of fundamental physics, can be expressed in terms of the
caesium hyperfine interval in atomic units and its variation may
be expressed in terms of the variations of $\alpha$ and $\mu_{\rm
Cs}/\mu_{\rm B}$. A constraint for the latter is presented in
Table~\ref{t:mic-99}.

\subsection{Constraints from Microwave Clocks}%xx.2
\label{susec:mwc-99}

A model-independent comparison of different HFS transitions is not
simple because their nonrelativistic contributions $f_{\rm NR}$
are not the same, but involve different magnetic moments. Applying
Eq.~(\ref{eqalp-99}) to experimental data, one can obtain
constraints on the relative variations of the magnetic moments of
Rb, Cs, and Yb (see Table~\ref{t:mic-99}).

\begin{table}
\caption{Model-independent laboratory constraints on the possible
time variations of natural constants. \label{t:mic-99}}
\begin{center}
\begin{tabular*}{246pt}{cr}
\hline
\hspace{25pt}$X$ & \multicolumn{1}{c}{\hspace{20pt}$\partial \ln X / \partial t$}  \\
\hline
\hspace{25pt}$\alpha$ & \hspace{20pt}$(-0.3\pm 2.0)\cdot 10^{-15}\,{\rm yr}^{-1}$\\
\hspace{25pt}$\{cR_\infty\}$ & \hspace{20pt}$(-2.1\pm 3.1)\cdot 10^{-15}\,{\rm yr}^{-1} $\\
\hspace{25pt}$\mu_{\rm Cs}/\mu_{\rm B}$ & \hspace{20pt}$(3.0\pm 6.8)\cdot 10^{-15}\,{\rm yr}^{-1} $\\
\hspace{25pt}$\mu_{\rm Rb}/\mu_{\rm Cs}$ & \hspace{20pt}$(-0.2\pm 1.2)\cdot 10^{-15}\,{\rm yr}^{-1} $\\
\hspace{25pt}$\mu_{\rm Yb}/\mu_{\rm Cs}$ & \hspace{20pt}$(3\pm 3)\cdot 10^{-14}\,{\rm yr}^{-1} $\\
\hline
\end{tabular*}
\end{center}
\end{table}

\subsection{Model-Dependent Constraints}%xx.2
\label{susec:mdc-99}

In order to gain information on constants more fundamental than
the nuclear magnetic moments, any further evaluation of the
experimental data should involve the Schmidt model, which is far
from perfect. Model-dependent constraints are summarized in
Table~\ref{t:mdc-99}.

\begin{table}
\caption{Model-dependent laboratory constraints on possible time
variations of fundamental constants. The uncertainties here do not
include uncertainties from the application of the Schmidt model.
\label{t:mdc-99}}
\begin{center}
\begin{tabular*}{246pt}{cr}
\hline
\hspace{25pt}$X$ & \multicolumn{1}{c}{\hspace{20pt}$\partial \ln X / \partial t$}  \\
\hline
\hspace{25pt}$m_{\rm e}/m_{\rm p}$ & \hspace{20pt}$(2.9\pm 6.2)\cdot 10^{-15}\,{\rm yr}^{-1} $\\
\hspace{25pt}$\mu_{\rm p}/\mu_{\rm e}$ & \hspace{20pt}$(2.9 \pm 5.8)\cdot 10^{-15}\,{\rm yr}^{-1} $\\
\hspace{25pt}$g_{\rm p}$ & \hspace{20pt}$(-0.1\pm 0.5)\cdot 10^{-15}\,{\rm yr}^{-1} $\\
\hspace{25pt}$g_{\rm n}$ & \hspace{20pt}$(3\pm 3)\cdot 10^{-14}\,{\rm yr}^{-1} $\\
\hline
\end{tabular*}
\end{center}
\end{table}

The nucleon $g$ factors, in their turn, depend on a dimensionless
fundamental constant $m_{\rm q}/\Lambda_{\rm QCD}$, where $m_{\rm
q}$ is the quark mass and $\Lambda_{\rm QCD}$ is the quantum
chromodynamic (QCD) scale. A study of this dependence may supply
us with deep insight into the possible variations of the more
fundamental properties of Nature (see Ref.\ \cite{flambaum04-99}
for details). This approach is promising, but its accuracy needs
to be better understood.

\section{Summary}%xx.5
\label{sec:sum-99}

The results collected in Tables~\ref{t:mic-99} and~\ref{t:mdc-99}
are competitive with data from other searches and have a more
reliable interpretation. The results from astrophysical searches
and the study of the samarium resonance from Oklo data claim
higher sensitivity (see, e.g., Ref.\ \cite{book-99}), however,
they are more difficult to interpret. We have, for example, not
assumed any hierarchy in variation rates or that some constants
stay fixed while others vary, as it is done in the study of the
position of the Oklo resonance. The evaluation presented here is
transparent, and any particular calculation or measurement can be
checked. In contrast, the astrophysical data show significant
results only after an intensive statistical evaluation.

The laboratory searches involving atomic clocks have definitely
shown progress and in a few years we expect an increase in the
accuracy of these clocks, an increase in the number of different
kinds of frequency standards (e.g., optical Sr, Sr$^+$, In$^+$
standards and a microwave Hg$^+$ standard are being tried now),
and indeed an increase in the time separation between accurate
experiments, since it is now typically only 2--3 years. An optical
clock based on a nuclear transition in Th-229 is also under
consideration \cite{th229-99}. Such a clock would offer different
sensitivity to systematic effects, as well as to variations of
different fundamental constants.

Laboratory searches are not necessarily limited by experiments
with metrological accuracy. An example of a high-sensitivity
search with a relatively low accuracy is the study of the
dysprosium atom for a determination of the splitting between the
$4f^{10}5d6s$ and $4f^{9}5d^26s$ states, which offers a great
sensitivity value of $\kappa\simeq5.7 \cdot 10^8$
\cite{dzuba03-99}.

Variations of constants on the cosmological time scale can be
expected but the magnitude, as well as other details, is unclear.
Because of a broad range of options there is a need for the
development of as many different searches as possible, and the
laboratory search for variations is an attractive opportunity to
open up a way that could lead to new physics.

\subsubsection*{Acknowledgments}

We are very grateful to our colleagues and to participants of
the ACFC-2003 meeting for useful and stimulating discussions.


\begin{thebibliography}{100}

\bibitem{here:Drake-99} W. E. Baylis and G. W. F. Drake, Chap.\ 1 in this Handbook.

\bibitem{codata-99}
P. J. Mohr and B. N. Taylor, Rev.\ Mod.\ Phys.\ (2004), to be
published.

\bibitem{sgk-99}
S. G. Karshenboim, Eprints physics/0306180 and physics/0311080, to
be published.

\bibitem{book-99}
{\em Astrophysics, Clocks and Fundamental Constants}, Lecture
Notes in Physics, edited by  S. G. Karshenboim and E. Peik
(Springer, Berlin, 2004), Vol.\ 648.

\bibitem{ramsey-99}
N. F. Ramsey, Rev.\ Mod.\ Phys.\ {\bf 62}, 541 (1990).

\bibitem{bauch-99}
A. Bauch, H. R. Telle, Rep.\ Prog.\ Phys.\ {\bf 65}, 789 (2002).

\bibitem{rb-99}
H. Marion, F. Pereira Dos Santos, M. Abgrall, S. Zhang, Y.
Sortais, S. Bize, I. Maksimovic, D. Calonico, J. Gruenert, C.
Mandache, P. Lemonde, G. Santarelli, Ph. Laurent, A. Clairon, and
C. Salomon, Phys.\ Rev.\ Lett.\ {\bf 90}, 150801 (2003).


\bibitem{paul-99}
W. Paul,  Rev.\ Mod.\ Phys.\ {\bf 62}, 531 (1990). See also: J.
Javanainen, Chap.\ 73 in this Handbook.

\bibitem{deh-99}
H. Dehmelt, IEEE Trans.\ Instrum.\ Meas.\ {\bf 31}, 83 (1982).

\bibitem{yb-99}
J. Stenger, C. Tamm, N. Haverkamp, S. Weyers, and H. R. Telle,
Opt.\ Lett.\ {\bf 26}, 1589 (2001).

\bibitem{hg-99}
T. Udem, S. A. Diddams, K. R. Vogel, C. W. Oates, E. A. Curtis, W.
D. Lee, W. M. Itano, R. E. Drullinger, J. C. Bergquist, and L.
Hollberg, Phys.\ Rev.\ Lett.\ {\bf 86}, 4996 (2001); S. Bize, S.
A. Diddams, U. Tanaka, C. E. Tanner, W. H. Oskay, R. E.
Drullinger, T. E. Parker, T. P. Heavner, S. R. Jefferts, L.
Hollberg, W. M. Itano, D. J. Wineland, and J. C. Bergquist, Phys.\
Rev.\ Lett.\ {\bf 90}, 150802 (2003).


\bibitem{ybhfs-99}
P. T. Fisk {\it et al.\/}, IEEE Trans.\ UFFC {\bf 44}, 344 (1997);
P. T. Fisk, Rep.\ Prog.\ Phys.\ {\bf 60}, 761 (1997); R. B.
Warrington, P. T. H. Fisk, M. J. Wouters, and M. A. Lawn, in {\em
Proceedings of the 6th Symposium Frequency Standards and
Metrology}, edited by P. Gill (World Scientific, 2002), p.\ 297.

\bibitem{borde-99}
C. J. Bord\'e, Phys.\ Lett.\ A {\bf 140}, 10 (1989).

\bibitem{ca-99}
G. Wilpers {\it et al.\/} , Phys.\ Rev.\ Lett.\ {\bf 89}, 230801
(2002); F. Riehle {\it et al.\/} in Ref.\ \cite{book-99}, p.\ 229.

\bibitem{katori-99}
H. Katori, M. Takamoto, V. G. Pal'chikov, and V. D. Ovsiannikov,
Phys.\ Rev.\ Lett.\ {\bf 91}, 173005 (2003).

\bibitem{h-99}
M. Niering, R. Holzwarth, J. Reichert, P. Pokasov, Th. Udem, M.
Weitz, T. W. H\"ansch, P. Lemonde, G. Santarelli, M. Abgrall, P.
Laurent, C. Salomon, and A. Clairon, Phys.\ Rev.\ Lett.\ {\bf 84},
5496 (2000); M. Fischer {\it et al.\/}, Phys.\ Rev.\ Lett.\ {\bf
92}, 230802 (2004).

\bibitem{chain-99}
T. Udem, J. Reichert, R. Holzwarth, S. Diddams, D. Jones, J. Ye,
S. Cundiff, T. W. H\"ansch, and J. Hall, in {\em The hydrogen
atom: Precision physics of simple atomic systems}, Lecture Notes
in Physics, edited by S. G. Karshenboim {\it et al.\/} (Springer,
Berlin, 2001), Vol.\ 570, p.\ 125.

\bibitem{ybnew-99}
E. Peik, B. Lipphardt, H. Schnatz, T. Schneider, Chr.\ Tamm, and
S. G. Karshenboim, physics/04021132.

\bibitem{here:Sapirstein-99}
J. Sapirstein, Chap.\ 28 in this Handbook; P. Mohr, Chap.\ 29 in
this Handbook.

\bibitem{savedoff-99}
M. P. Savedoff, Nature {\bf 178}, 688 (1956).

\bibitem{thompson-99}
R. I. Thompson, Astrophys.\ Lett.\ {\bf 16}, 3 (1975).

\bibitem{varshalovich-99}
D. A. Varshalovich, A. V. Ivanchik, A. V. Orlov, A. Y. Potekhin,
and P. Petitjean, in {\em Precision Physics of Simple Atomic
Systems}. Lecture Notes in Physics, edited by S. G. Karshenboim
and V. B. Smirnov (Springer-Verlag, Berlin, 2003), Vol.\ 627, p.\
199.

\bibitem{Karshenboim-99}
S. G. Karshenboim, Can.\ J. Phys.\ {\bf 78}, 639 (2000).

\bibitem{flambaum04-99}
V. V. Flambaum, physics/0309107; V. V. Flambaum, L.B. Leinweber,
A.W. Thomas, and R.D. Young, Phys.\ Rev.\ D {\bf 69}, 115006
(2004).

\bibitem{prestage-99}
J. D. Prestage, R. L. Tjoelker, and L. Maleki, Phys.\ Rev.\ Lett.\
{\bf 74}, 3511 (1995).

\bibitem{dzuba1-99}
V. A. Dzuba, V. V. Flambaum, and J. K. Webb, Phys.\ Rev.\ Lett.\
{\bf 82}, 888 (1999); Phys.\ Rev.\ A {\bf 59}, 230 (1999).

\bibitem{dzuba01-99}
V. A. Dzuba, V. V. Flambaum, Phys.\ Rev.\ A {\bf 61}, 034502
(2001).

\bibitem{dzuba03-99}
V. A. Dzuba, V. V. Flambaum, M.V. Marchenko, Phys.\ Rev.\ A {\bf
68}, 022506 (2003).

\bibitem{quinn-99} T. J. Quinn, Metrologia {\bf 40}, 103 (2003).

\bibitem{th229-99}
E. Peik and Chr.\ Tamm, Europhys.\ Lett.\ {\bf 61}, 181 (2003).

\end{thebibliography}
\end{document}